\documentclass[12pt]{article}

\usepackage{scicite}

\usepackage{times}
\usepackage{graphicx}
\usepackage[normalem]{ulem}
\usepackage{color}

\newenvironment{sciabstract}{%
\begin{quote} \bf}
{\end{quote}}

\newcounter{lastnote}

\title{Binary superlattice design by controlling DNA-mediated interactions} 

\author
{Minseok Song, Yajun Ding, Hasan Zerze, Mark A. Snyder, Jeetain Mittal$^{\ast}$\\
\\
\normalsize{Department of Chemical and Biomolecular Engineering, Lehigh University,}\\
\normalsize{111 Research Drive, Iacooca Hall, Bethlehem, PA 18015, USA}\\
\normalsize{$^\ast$To whom correspondence should be addressed; E-mail:  jeetain@lehigh.edu.}
}

\date{}

\begin{document} 


\baselineskip24pt


\maketitle


\begin{sciabstract}
Most binary superlattices created using DNA functionalization or other approaches rely on particle size differences to achieve compositional order and structural diversity. Here we study two-dimensional (2D) assembly of DNA-functionalized micron-sized particles (DFPs), and employ a strategy that leverages the tunable disparity in interparticle interactions, and thus enthalpic driving forces, to open new avenues for design of binary superlattices that do not rely on the ability to tune particle size (i.e., entropic driving forces). Our strategy employs tailored blends of complementary strands of ssDNA to control interparticle interactions between micron-sized silica particles in a binary mixture to create compositionally diverse 2D lattices. 
We show that the particle arrangement can be further controlled by changing the stoichiometry of the binary mixture in certain cases. With this approach, we demonstrate the ability to program the particle assembly into square, pentagonal, and hexagonal lattices. In addition, different particle types can be compositionally ordered in square checkerboard and hexagonal -- alternating string,  honeycomb, and Kagome arrangements. 
\end{sciabstract}



The field of DNA-mediated particle assembly has undergone remarkable progress over recent years\cite{jones2015programmable}, owing, at least in part, to its potential as a powerful platform for rational, bottom-up design and engineering of complex materials, and motivated by recent successful translations into applications as diverse as sensing\cite{Barnaby2015}, photonics\cite{Ross2015}, and catalysis. The growing number of synthetic pathways and design strategies to fabricate DNA-functionalized particles (DFPs) has led to the development of a diverse palette of tailorable building blocks from which to choose, comprised of particles of a wide range of inorganic to organic compositions, a near continuum of particle sizes spanning nanometers to micrometers, precise DNA sequence control and thus tailorable hybridization, diverse chemistries for DNA grafting/association, and fine tunability of the grafting density.\cite{Kim2006,zhang2013general,Oh2015} 

Accompanying this expanding diversity of building blocks has been a parallel development of specific to generalized design principles that have begun to link molecular-scale DFP function with mechanisms of assembly and the resulting uni- or multi-modal crystalline structures. To this end, the growing combination of theory, simulations, and experiments, has helped to overcome some of the challenges in the field. For example, re-entrant melting strategies\cite{Angioletti2012, Rogers2015} have been successfully developed to alleviate the very narrow temperature ranges for efficient crystallization of DFPs. 

The most common route to induce attraction between DFPs, and thus program their assembly, leverages the direct or indirect (i.e., with additional DNA linker strand) hybridization of complementary DNA strands tethered separately to two types of particles. Under suitable conditions in such systems, particles with complementary DNA functionality (i.e., `unlike' particles) form attractive contacts among multiple strands of hybridizable DNA, whereas particles bearing the same DNA functionality (i.e., `like' particles) typically interact via repulsive, non-hybridizable DNA-mediated steric interactions. By tailoring DFP properties such as particle size, DNA sequence, DNA strand length, and strand grafting density, three-dimensional (3D) assembly of a diversity of stoichiometric and symmetry structures including CsCl, AlB$_2$, Cr$_3$Si, Cs$_6$C$_{60}$, NaTl, and others has been demonstrated for nanoparticle systems~\cite{Nykypanchuk2008, Macfarlane2011, Li2012,Auyeung2014,Cigler2010}. 

In these cases, assembly is ultimately tailored by controlling properties of particles that are each functionalized with only a single type of ssDNA. As an alternative, mixtures of two complementary DNA strands in a desired ratio can be grafted to particles as a means for tuning pair-interactions. This, so-called multiflavoring scheme was proposed to selectively program 3D crystallization of DFP into body-centered cubic (BCC) or closed-packed structures (CP)~\cite{Scarlett2011,Casey2012}. Furthermore, Zhang and coauthors showed that such muliflavored DFPs can also be utilized as reconfigurable systems that are capable of dynamically transforming into BCC and CP lattices with the addition of suitable DNA sequences~\cite{Zhang2015}. This suggests that control over interparticle interactions may serve as a new handle for dramatically expanding the structural diversity of crystalline assemblies and for realizing novel smart and adaptable materials.

\begin{figure}[h!]
\center
\includegraphics[width = 6.5 in]{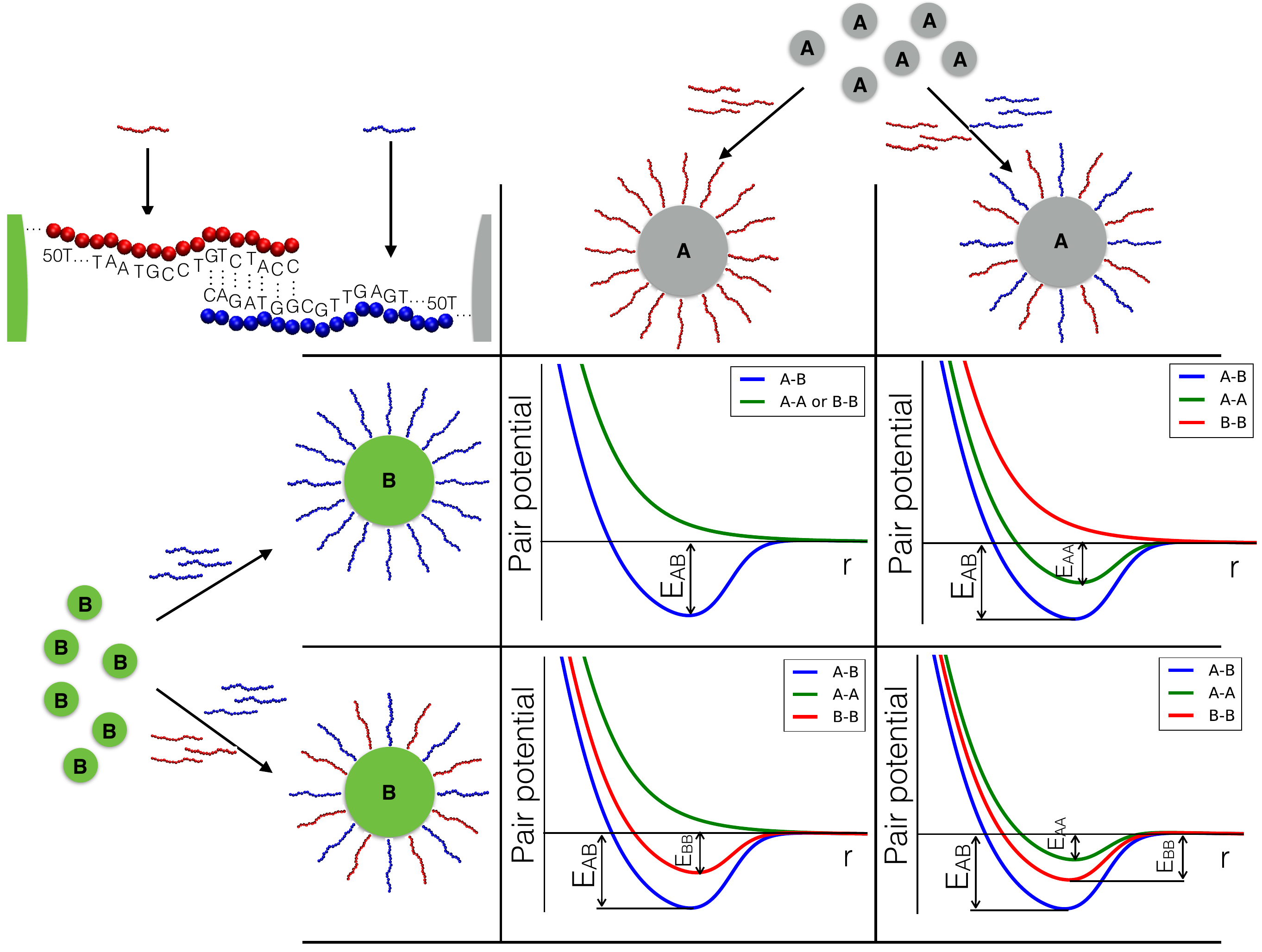}
    \caption{Multiflavoring scheme. Programming the particle interactions through functionalizing with a blend of conjugate DNA strands, $\alpha$DNA (depicted as red strands) and $\beta$DNA (depicted as blue strands) at desired ratios. The three resulting pair interactions ($E_{AA}, E_{BB}, E_{AB}=E_{BA}$) can be independently controlled as shown in the quadrant of panels. 
    }
    \label{fig1}
\end{figure}

Although nanoparticle-based DNA-mediated assembly enables realization of a diversity of crystalline structures, the assembly, especially in two-dimensions, of micron-sized DFPs has been slow to come. The appeal of micron-sized particle assemblies, and the motivation of efforts to overcome this barrier, is driven from an applications perspective by their desirable optical properties~\cite{Park2014}, and, from a fundamental science perspective, by the ability to employ simpler optical monitoring for direct mechanistic insight into DFP crystallization~\cite{Gasser2001}. Micron-sized DFPs possess relatively short-range interactions compared to their sizes, and extremely narrow melting transition ranges (1-2$^{o}$C), which leads to tedious annealing protocols with many days of incubation for successful crystallization ~\cite{Biancaniello2005,Kim2006,Dreyfus2010,DiMichele2013,Wang2015,Wang2015i}. Recently, it was shown that DNA strand displacement can be used to widen the melting transition range and to shorten the effective crystallization time~\cite{Rogers2015}. Previous studies on the 2D assembly of DNA-tethered colloids have sought to better understand their association/dissociation transitions~\cite{Dreyfus2009,Dreyfus2010,song2016effect}, and have relied on surface-mediated templating of DFP monolayers~\cite{Hartmann2002,Zou2005,Puchner2008,Shyr2008}, 
but the direct formation of DNA-mediated binary 2D superlattices remains a serious challenge.

The current paradigm for binary superlattice formation primarily leverages entropic packing effects to generate structures with tailored lattice symmetries. This represents a natural extension of extensively characterized binary hard sphere-like mixtures~\cite{kung2014template} as well as ionic crystals~\cite{leunissen2005ionic}. 
Here, we present an enthalpic design strategy, which takes advantage of tuning the strength of interparticle attractions between equally-sized, but distinctly labeled particles in a binary (A, B)  mixture. Interactions between like (AA and BB) and unlike (AB) particle pairs are independently controlled to guide their assembly into compositionally ordered lattices. Extensive computer simulations enable comprehensive investigation of this binary assembly, leading to the prediction of compositionally ordered two-dimensional (2D) structures such as square, alternating string, honeycomb, Kagome and square Kagome that we have verified experimentally by leveraging a previously proposed multiflavoring approach~\cite{Casey2012, Scarlett2011} to tune interparticle interactions among micron-sized DFPs by judicious blending of complementary ssDNA functionality as shown in Figure~\ref{fig1}.

First, we have performed molecular dynamics (MD) simulations to investigate the two-dimensional (2D) assembly of binary micron-sized DFPs over a wide range of possible combinations in their interaction strengths (see Figure 1 for details). 
For simplicity in presenting and interpreting results, we keep unlike pair interaction strength, $E_{\mathrm {AB}}$, fixed and vary like pair interaction strengths, $E_{\mathrm {AA}}$ and $E_{\mathrm {BB}}$, independently. All of the results are presented as a function of normalized pair energies, $E_{\mathrm {AA}}$/$E_{\mathrm {AB}}$ and $E_{\mathrm {BB}}$/$E_{\mathrm {AB}}$. 

To model the short-range interparticle interactions mediated by DNA interactions between micron-sized colloidal particles, we use a suitable functional form (see Supplementary Information (SI) and Fig. S1 for details) that captures important details of the underlying repulsive (particle-particle and osmotic repulsion due to overlap between DNA chains) as well as attractive (due to DNA hybridization) parts of the potential. Similar pair potentials between DFPs and other complex systems have been used in the past to successfully capture their self-assembly properties~\cite{rabideau2007computational,tkachenko2002morphological,martinez2011design, Auyeung2014,mahynski2015grafted}. 

To identify suitable conditions for superlattice formation, we conduct simulations over a large range of temperatures and identify putative melting transition temperatures based on the changes in potential energy as a function of temperature (Fig. S2). At temperatures close to the melting transition, sufficiently large crystalline assemblies are observed that can be used for further analysis. We note that most of the simulation results reported in this paper are based on systems at relatively low number density ($\rho$ = 0.10) to identify design parameters suitable for enthalpically-driven assembly of binary superlattices.  To identify different lattice symmetries and underlying compositional order in the binary superlattices that are formed, we use three complementary methods for which details are provided in the SI: (i) nearest neighbor analysis (NNA) (SI Figs. S3), (ii) common neighbor analysis (CNA)~\cite{Stukowski2012} (SI Fig. S4), and (iii) visual inspection using visual molecular dynamics (VMD)~\cite{Humphrey1996}.

\begin{figure}[h!]
\center
\includegraphics[width = 6.5 in]{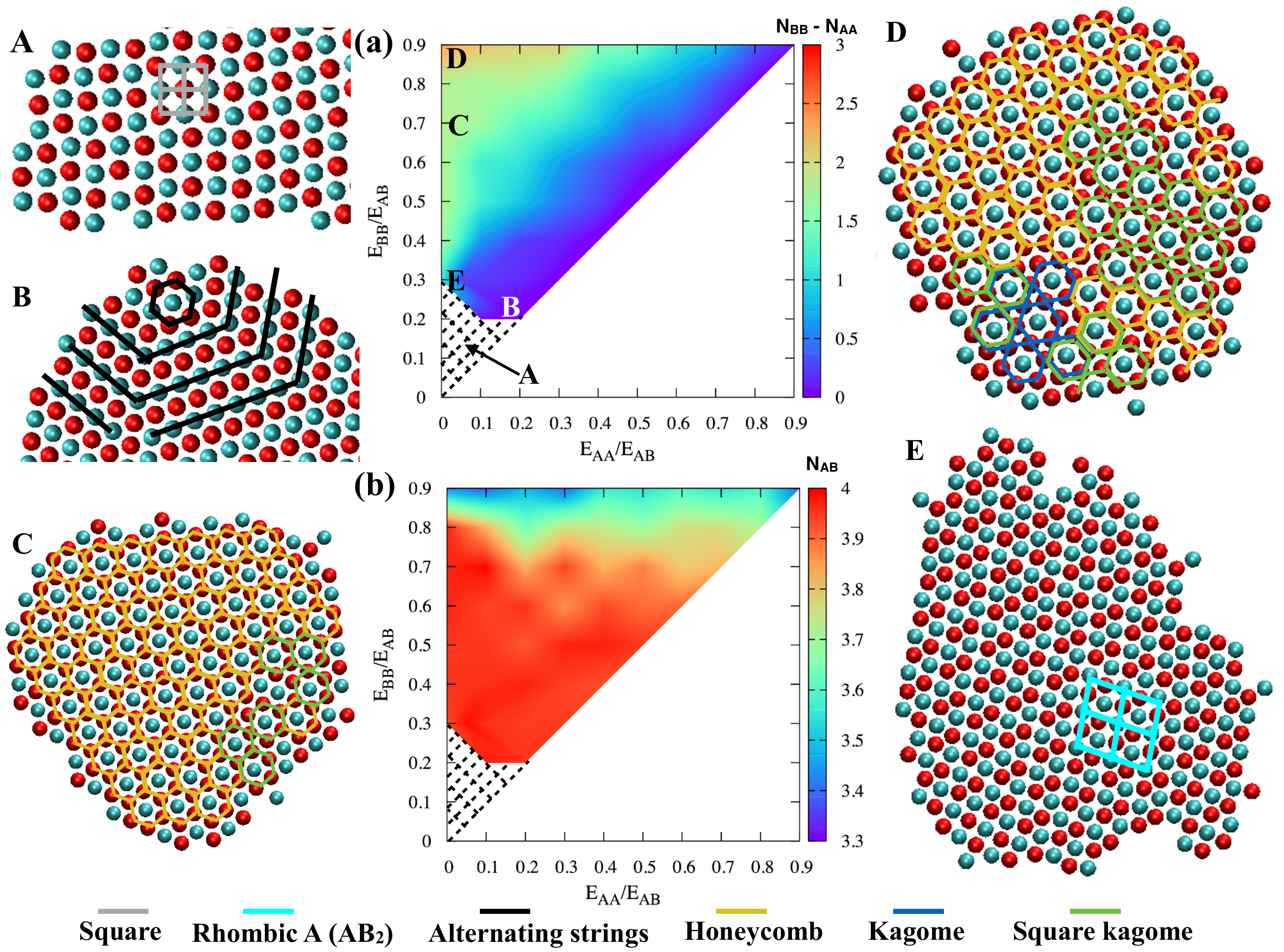}
    \caption{Order diagram showing a variety of 2D binary crystals controlled as a function of interparticle binding strengths. The color coded lines drawn on the crystal images (A-E) are guides to the eye for illustrating the corresponding lattice symmetries. Color maps show the identification of the crystals using \textbf{(a)} $N_{BB}-N_{AA}$ and \textbf{(b)} $N_{AB}$ as the order parameter.}
    \label{fig2}
\end{figure}

Extensive results from MD simulations are summarized in Figure~\ref{fig2} as order diagrams depicting 2D binary crystal symmetries resulting from tuned interparticle interactions as a function of $E_{\mathrm {AA}}$/$E_{\mathrm {AB}}$, $E_{\mathrm {BB}}$/$E_{\mathrm {AB}}$.  
NNA (Table 1) enabled interpretation of structural symmetries from computational results, which was confirmed by CNA and visual inspection. These data show that various lattice symmetries (square, hexagonal, rhombic) can be obtained by simply changing $E_{\mathrm {AA}}$/$E_{\mathrm {AB}}$ and/or $E_{\mathrm {BB}}$/$E_{\mathrm {AB}}$, as depicted in the snapshots shown in Figure~\ref{fig2}(A-E), of representative binary crystal structures. For low values of like pair energies, one can expect to form non-close-packed square lattice structures as quantified by NNA and depicted as the hatched area in Fig.~\ref{fig2}(a-b). Such structures are otherwise difficult to obtain without introducing additional factors such as surface templating~\cite{ferraro2014graphoepitaxy}. 
Previous investigations of 3D assembly of DFPs, where like pair interactions are commonly purely repulsive ($E_{\mathrm {AA}}$/$E_{\mathrm {AB}}$ = $E_{\mathrm {BB}}$/$E_{\mathrm {AB}}$ = 0), have also found that non-closed-packed BCC structures are stabilized~\cite{Li2012}.  
As shown in Fig.~\ref{fig2}a, the predicted two-dimensional square lattices display a high degree of compositional order in terms of the coordinated arrangement of A and B particles. Further increasing like-pair energies leads to the formation of close-packed hexagonal lattices, with an interesting rhombic structure formed for the case specified as E in Fig.~\ref{fig2}, and identified only visually. The rhombic structure is of the A(AB$_2$) type in this case. 

\begin{table}[h!]
\begin{center}
\begin{tabular}{ c c c c }
\hline
 {\bf Crystal} & {\bf N$_{\mathrm {AB}}$} & {\bf N$_{\mathrm {AA}}$} & {\bf N$_{\mathrm {BB}}$} \\
 \hline
 Square & 4 & 0 & 0 \\  
 Alternating string & 4 & 1 & 1 \\
 Honeycomb & 4 & 0 & 2 \\
 Kagome & 3 & 0 & 3 \\
 Square Kagome & 3 & 0 & 3 \\
\hline
\end{tabular}
\caption{Reference nearest neighbor counts for perfect two-dimensional binary superlattices, where N$_{\mathrm {AB}}$ is the number of unlike contacts and N$_{\mathrm {AA}}$ or N$_{\mathrm {BB}}$ is the number of like contacts.}
\label{table1}
\end{center}
\end{table}

For hexagonally packed structures, we identify conditions for many different binary superlattices, namely alternating strings, honeycomb, Kagome, and square Kagome (Fig.~\ref{fig2}(B-D)). 
In a small region of moderate like pair energies, hexagonal structures are compositionally ordered as alternating strings (AS) (Fig.~\ref{fig2}B). Detailed analysis of MD trajectories (not shown here) suggest the formation of AS lattices due to the transformation of nuclei, initially grown as square into hexagonal lattices (see SI movie). Such diffusionless transformations have previously been observed in the context of DFPs in 3D systems~\cite{casey2012driving}, but the underlying mechanism is still not fully resolved~\cite{jenkins2014hydrodynamics}. We are currently investigating this issue in more detail, which will be discussed in a future publication.  

With increasing like pair energies along the diagonal axis ($E_{\mathrm {AA}}$/$E_{\mathrm {AB}}$ = $E_{\mathrm {BB}}$/$E_{\mathrm {AB}}$), hexagonal structures become compositionally disordered, which is expected as the energetic difference in interparticle interactions becomes smaller. Interestingly, when $E_{\mathrm {AA}}$/$E_{\mathrm {AB}} > E_{\mathrm {BB}}$/$E_{\mathrm {AB}}$, hexagonal structures can be found in many compositionally ordered arrangements such as honeycomb, Kagome and square Kagome lattices. As evident in Fig.~\ref{fig2}a, honeycomb arrangement is quite prevalent in a large region of the order diagram. 
Kagome and square Kagome arrangements are observed when $E_{\mathrm {AA}}$/$E_{\mathrm {AB}} \ll E_{\mathrm {BB}}$/$E_{\mathrm {AB}}$ $\approx$ 1. In these cases, contacts between unlike pairs (AB) are only slightly more favored energetically than the like pair (BB), which gives rise to the formation of binary superlattices with non-equimolar A:B stoichiometries (e.g., 1:2 for honeycomb and 1:3 for Kagome).
These simulation results ultimately demonstrate that simple tuning of interparticle attractions can lead to the formation of a diverse array of binary superlattice structures in 2D systems 
without requiring blending of particles of different sizes, application of external fields, or introduction of surface structuring or other external factors. In order to test the validity of these results as possible guidelines for rational design of binary superlattices, we have carried out complementary experimental studies of 2D particle assembly wherein blending of complementary ssDNA particle functionality was used as a means for controlling interparticle interactions and mediating assembly of distinct fluorescently labeled particle populations. 

Specifically, like-sized (1.5 $\mu$m) red (A) and green (B) fluorescent silica particles were separately functionalized with pure or blended complementary single-stranded $\alpha$DNA and $\beta$DNA using sequential silanization and cyanuric chloride chemistries as reported previously\cite{song2016effect}. Their assembly was followed via optical and fluorescence microscopy. While the blending ratio, $\gamma_i=\alpha$DNA/($\alpha$DNA+$\beta$DNA), was tuned independently between pure $\alpha$DNA and pure $\beta$DNA for B-type particles ($\gamma_B=0-1$), A-type particles were functionalized with pure $\alpha$DNA ($\gamma_A=1$) as a way to experimentally traverse the most relevant parameter space mapped out by our computations ($E_{AA}/E_{AB}=0, E_{BB}/E_{AB}=[0,1]$). This allowed for programming of attractive interaction strengths between like ($E_{\mathrm {AA}}$, $E_{\mathrm {BB}}$) and unlike ($E_{\mathrm {AB}}$) particles (Fig. \ref{fig1}). 
Our analysis suggests that increases in the relative loading of $\alpha$DNA on the B-type particles should lead to a monotonic decrease in the strength of attractive interactions between unlike particles ($E_{AB}=E_{BA}$) while the strength of attraction between B-type particles ($E_{BB}$) should monotonically increase. Such tunability of interparticle interaction strengths should ultimately result in a scenario where interactions between one set of like particles (B) become equivalent to or even exceed interactions between unlike particles. 

As an initial confirmation of the tunability of the interparticle attraction, we measured the melting transition of both unary (only B-type) and binary (both A- and B-type) DFPs at different $\gamma_B$ while holding $\gamma_A=1$. With increasing $\gamma_B$, the melting temperature of the unary mixture (Supplementary Fig. S5) shifts to lower temperatures, whereas that of the binary mixture shifts to higher temperatures. These data clearly demonstrate that with increasing $\gamma_B$ the strength of interactions between B-type particles systematically decreases while the strength of interactions between unlike particles systematically increases.
  

\begin{figure}[h!]
\center
\includegraphics[width = 3.25 in]{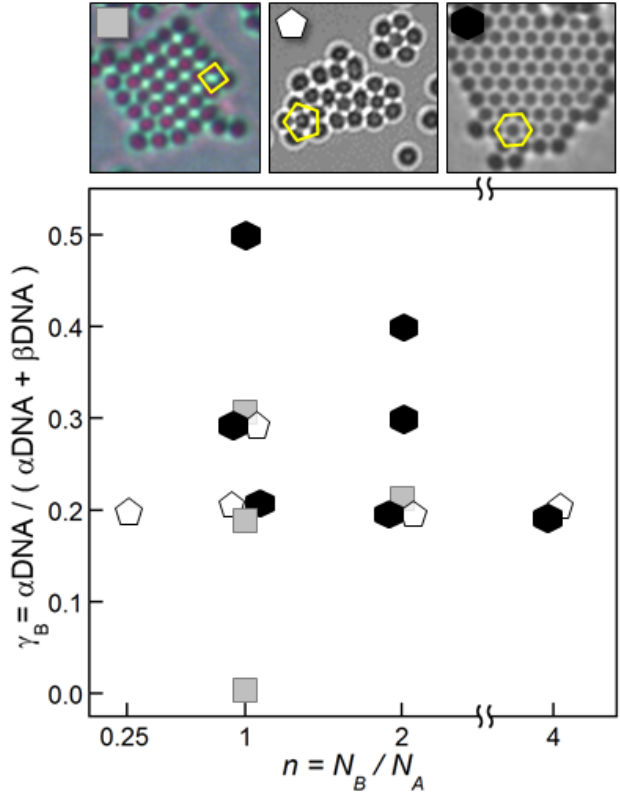}
    \caption{Identification of structural order within 2D DFP lattices derived from solutions of different particle number ratios, $n$, and DFPs with different DNA blending ratios, $\gamma_B$ ($\gamma_A=1$). The phase diagram shows the characteristic structures observed under the specified conditions, including square (grey squares), pentagonal (white pentagons), and and hexagonal (black hexagons) lattices for which the lattice symmetry group or mixtures of groups are depicted with concomitantly shaped symbols. Particle assembly images show representative snapshots of square, pentagonal, and hexagonal lattices.
    }
    \label{fig3}
\end{figure}

Beyond determination of the melting temperature, experimental characterization focused on identification of crystal structures emerging upon self-assembly of the DFPs with systematically controlled ssDNA blending ratios $\gamma_B$ ($\gamma_A=1$) and various particle number ratios, $n=N_B/N_A$, where $N_{\mathrm A}$ and $N_{\mathrm B}$ are the number of A and B particles in solution, respectively. The tuning of of $n$ was motivated by previous studies wherein  it was shown that solution stoichiometry can have a significant impact on the formation of three-dimensional binary superlattices formed by DNA-mediated interactions~\cite{vo2015stoichiometric}.
Changing the solution stoichiometry away from 1:1 (A:B) tends to favor structures with symmetries that closely resemble the stoichiometry of the solution itself. For example AB$_2$ and AB$_3$ solid stoichiometric structures may be favored over AB for 1:2 and 1:3 binary mixtures, respectively. 

Here, we have assembled 2D structures from solutions of A and B DFPs of specified $n$, varied from 0.25 to 4. Under these conditions, the ssDNA blending ratio for B particles, $\gamma_B$, was systematically varied while A particles were functionalized only with $\alpha$DNA ($\gamma_A=1$). After the completion of thermal annealing, images of the assembled DFPs were collected at room temperature using an inverted optical microscope. Local structural analysis and calculation of pair correlations for broader fields of particles was carried out to identify the formation of specific 2D crystal structures, to determine their symmetry, and, by way of combined analysis of bright field and fluorescence images, to evaluate the compositional (i.e., A-B) order of the structures.   

The structural order of 2D assemblies of A and B DFPs as a function of $\gamma_{\mathrm B}$ 
and $n$ 
are summarized in Fig. \ref{fig3}. The symbol shapes reflect the observed crystalline lattices ranging among square, pentagonal, and hexagonal symmetries as well as mixtures thereof. As $\gamma_{\mathrm B}$ 
is increased with $n$ held constant (e.g., $n=1$), we observe the formation of square lattice configurations at the lowest blending ratios considered ($E_{\mathrm {BB}}$ much less than $E_{\mathrm {AB}}$), the onset of minority hexagonal structures at higher blending ratios, eventual transition to majority hexagonal structures, and final dominance by hexagonal lattices at the highest blending ratios ($E_{\mathrm {BB}}$ comparable to $E_{\mathrm {AB}}$). These results are quite consistent with the computationally-derived order diagram presented in Fig.~\ref{fig2} with the exception of the fact that we also observe the formation of a minority phase of pentagonal lattices for intermediate blending ratios, which was not expected from the simulation results. The exact reasoning behind the appearance of such lattices in our experimental systems is somewhat unclear, but this may derive from the difference in local concentration of A and B particles relative to the prescribed bulk solution stoichiometry itself. 

As interparticle interactions between A particles are purely repulsive ($\gamma_{\mathrm A}=1$),  
increasing the amount of B particles in the solution may favor the formation of lattices with a higher number of BB contacts. We generally observe such a change with increasing $n=N_{B}/N_{A}$. For example, as $n$ is increased while holding $\gamma=0.4$, the dominance of square lattice structures at the lowest $n$ studied gives way to structures that are largely hexagonal. Here, again, we have identified persistent minority petagonal structures for $n$ greater than or equal to 1. In addition, we find that underpopulation of B particles in the solution ($n=0.25$) can lead to the formation of pure pentagonal structures. This is likely due to the higher local availability of B particles with respect to the solution stoichiometry, but inadequacy at the same time to stabilize hexagonal lattice structures. Entropic packing effects undoubtedly contribute to this observation as well, but these are difficult to decouple from the role of stoichiometry in the lattice selection.  

\begin{figure*}[h!]
\center
\includegraphics[width = 6.5 in]{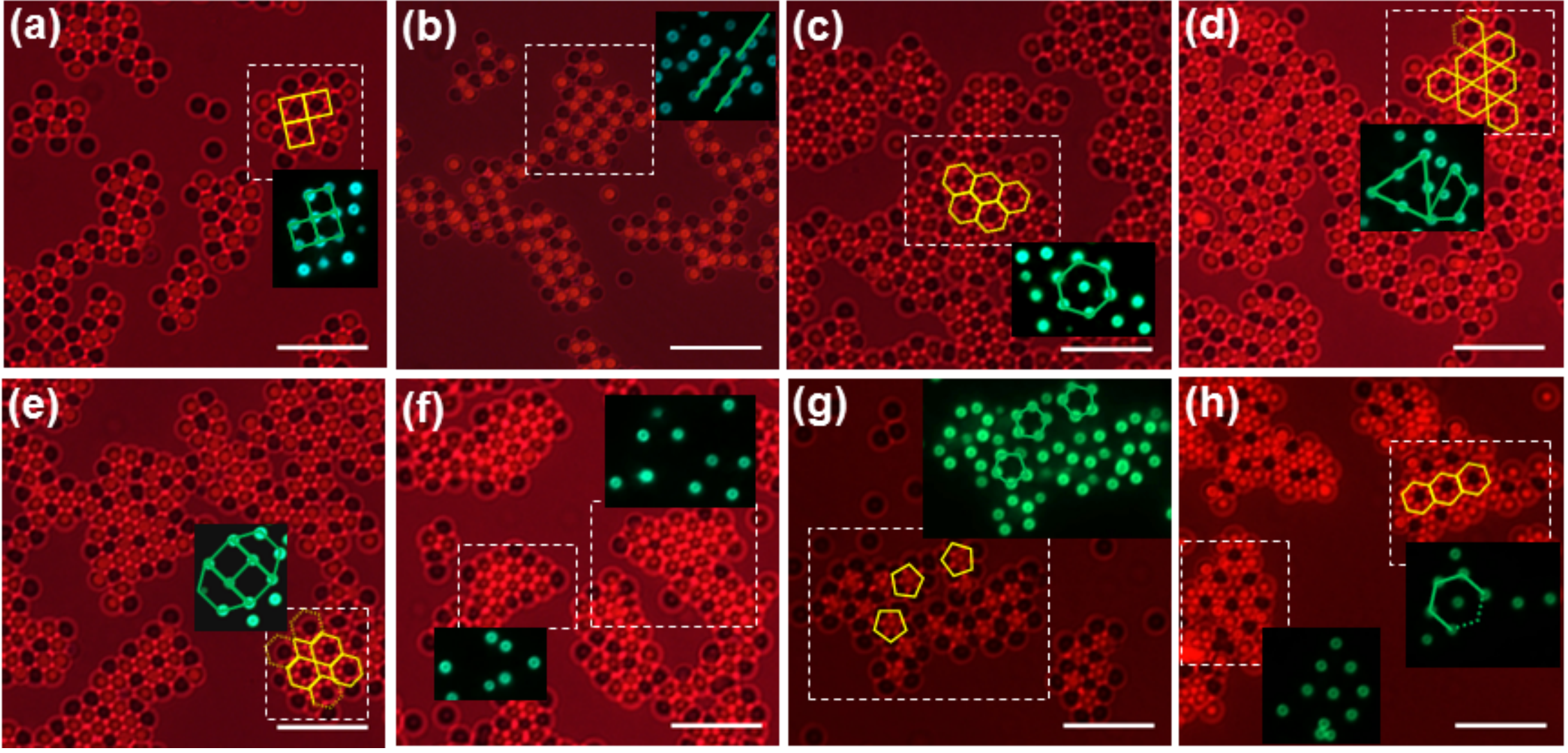}
    \caption{\footnotesize Identification of compositional order in representative fluorescence images of 2D DFP crystals comprised of A (dark) and B (bright) particles. Insets show B-specific fluorescence images (i.e., green channel) associated with specified regions (dashed boxes) of the crystal structure comprised of  
    \textbf{(a)} square lattices ($\gamma_B=0.2$, $n=1$), 
    \textbf{(b)} alternating strings ($\gamma_B=0.2$, $n=1$),
    \textbf{(c)} honeycomb structures ($\gamma_B=0.3$, $n=2$),
    \textbf{(d)} Kagome and honeycomb structures ($\gamma_B=0.3$, $n=2$),  
    \textbf{(e)} square Kagome and honeycomb structures ($\gamma_B=0.3$, $n=2$),
    \textbf{(f)} demixed A and B particles ($\gamma_B=0.4$, $n=2$)
    \textbf{(g)} pentagonal structures ($\gamma_B=0.2$, $n=0.25$), and
    \textbf{(h)} honeycomb structures ($\gamma_B=0.2$, $n=4$). Scale bars in each image represent 10$\mu$m.}

   \label{fig4}
\end{figure*}

In short, our experimental data validate the computational findings, demonstrating that the formation of specific lattice structures can be tailored simply by the combined design of the strength of interparticle interactions through control of the ssDNA blending ratio, $\gamma_B$, and the particle number ratio, $n$, as opposed to surface patterning or other external factors. In addition to structural order, however, we are interested in assessing how well compositional order can also be controlled. Specifically, we employ multi-channel red and green fluorescence imaging to differentiate A and B particles, respectively, within DFP assemblies to determine the achievable compositional coordination and symmetry (Fig. \ref{fig4}). 
We find generally that compositional order can, indeed, be tuned by controlling the blending ratio $\gamma_i$, and thus interparticle interaction strength as wel as the particle stoichiometry, $n$. Moreover, unlike the simulations, simultaneous control over stoichiometry, $n$, and interparticle attraction strength may be critical for realizing higher symmetry lattices such as Kagome and square Kagome, and for controlling polymorphism. 

Fig.~\ref{fig4} depicts the diversity of achievable compositional order and symmetry accompanying the previously discussed structural tunabilty.
For low $\gamma_B$ ($\gamma_A=1$), particles preferentially organize into square lattices. In those structures, A and B particles are compositionally well-ordered as shown in (Fig. \ref{fig4}a) for $n=1$. This is fully consistent with the particle arrangement observed in the simulations, as very few compositional defects are seen for square lattices. 

As shown in Fig. 3, one can observe hexagonal lattices by either increasing the like-pair energies or $n$ (e.g., Fig. \ref{fig4}h). If we hold $n$ constant at 2, and change the blending ratio (i.e., to increase like-pair attraction between B particles), we observe compositional ordering into alternating string structures (Fig. \ref{fig4}b), honeycomb structures (Fig. \ref{fig4}c), Kagome structures (Fig. \ref{fig4}d), and square Kagome structures (Fig. \ref{fig4}e) as well as segregated B particles (Fig. \ref{fig4}f). In full consistency with our simulation results (Fig. \ref{fig2}), we have observed the coexistence of honeycomb and Kagome structures for the same blending ratio ($\gamma_B=0.3$). 
As previously discussed, we also observe the emergence of pure pentagonal lattices at the lowest $n$ studied here ($n=0.25$). Fluorescence images provide insight into the compositional order of these lattices, which is shown in Fig. \ref{fig4}g. As expected, each A particle is surrounded by 5 B particles, thereby maximizing A-B and B-B coordination.

Taken together with the computational analysis presented before, the validating insight provided by laboratory assembly of DFPs reveals that DNA-mediated particle assembly can enable the formation of 2D binary crystalline structures with a diversity of both structural and compositional order and symmetry through enthalpically rather than entropically controlled processes.
Specifically, without any template or tuning of the size of multi-modal DFP mixtures, square, hexagonal lattice and pentagonal aggregates with alternating strand, honeycomb, Kagome and square Kagome binary compositional symmetries emerge as a result of the controlled interplay between attractive interparticle interactions and tailored solution stoichiometry. This enthalpic handle for controlling structural and compositional diversity enhances the control afforded by entropic effects for realizing material complexity and offers potentially new routes to the design and synthesis of material function, hierarchically intercalated structures, or the sacrificial templating of hierarchically porous materials.



\section*{Acknowledgments}
We thank Professor Ben Rogers (Brandeis University) for critical input with the experimental setup and for helpful discussions. This work was supported by the U.S. Department of Energy, Office of Basic Energy Science, Division of Material Sciences and Engineering under Award (DE-SC0013979). This research used resources of the National Energy Research Scientific Computing Center, a DOE Office of Science User Facility supported under Contract No. DE-AC02-05CH11231.
Use of the high-performance computing capabilities of the Extreme Science and Engineering Discovery Environment (XSEDE), which is
supported by the National Science Foundation, project no. TG-MCB120014, is also gratefully acknowledged.

\section*{Methods}
\subsection*{Particle functionalization}
Green (B) and red (A) fluorescent 1.5-$\mu$m-diameter silica particles (micromod Partikeltechnologie GmbH, Germany) were separately functionalized with blends of complementary single-stranded 5$^\prime$-primary amine-modified DNA (ssDNA, Integrated DNA Technologies, Inc), using silanization and cyanuric chloride chemistries as reported previously.\cite{Steinberg2004} The specific sequences of the complementary ssDNA, which we refer to as $\alpha$DNA and $\beta$DNA, were 5$^\prime$-NH$_2$-(CH$_{2}$)$_{6}$-(T)$_{50}$-TAATGCCTGTCTACC-3$^\prime$ and 5$^\prime$-NH$_2$-(CH$_2$)$_{6}$-(T)$_{50}$-TGAGTTGCGGTAGAC-3$^\prime$, respectively.

Briefly, silica particles were first functionalized with (3-aminopropyl) triethoxysilane (APTES, Acros Organics) in ethanol. Following solvent exchange, cyanuric chloride (CCl, Sigma-Aldrich) was reacted with the primary amine group of APTES on the particles in a solution of acetonitrile and N,N-diisopropylethylamine (Sigma-Aldrich). The covalent attachment of amine-modified oligonucleotides to the surface of CCl-functionalized silica was carried out in salty borate buffer (pH 8.5), with the $\alpha$DNA:$\beta$DNA molar composition of the functionalization solution tuned to match the desired blending ratios, $\gamma_i=\alpha$DNA/($\alpha$DNA+$\beta$DNA), of $\alpha$DNA and $\beta$DNA on the red ($\gamma_A$) and green ($\gamma_B$) particle surface. After completion of the reaction, DNA-functionalized particles were washed and redispersed in TE buffer (pH 8.0), consisting of 10 mM Tris-HCl and 1 mM EDTA, with 100 mM NaCl. The surface density of DNA was estimated by following the same functionalization procedure, but with non-fluorescent silica particles and fluorescently labeled ssDNA (5$^\prime$-NH$_{2}$-(CH$_{2}$)$_{6}$-TTTTTTATGTATCAAGGT-Cy5-3$^\prime$). Fluorimeter measurements estimate the total ssDNA density at ca. 38,000 strands/$\mu$m$^2$. Since the fluorescent DNA bears a different sequence and lower number of base pairs (18 bps) compared to the nominal DNA employed in this study (65 bps), we anticipate this estimated density may serve as an upper bound for the actual density of the 65-bp ssDNA on the silica surface.
\subsection*{Sample preparation} Suspensions containing mixtures of A and B ssDNA-functionalized particles (DFPs) at desired particle number ratios, $n=N_A/N_B$, where $N_A$ and $N_B$ are the number density of A and B particles, respectively, were prepared from stock suspensions of each particle type by washing and redispersion in 100 mM NaCl TE buffer. 0.1 wt$\%$ Pluronic F88 (BASF) was included to protect against non-specific binding between the particles themselves as well as between the particles and the glass surface of the coverslip chamber employed for imaging of particle assembly. The coverslip microchamber was comprised of two plasma-treated coverslips bonded together on three sides by a melted and solidified parafilm seal ($\sim$ 250 $\mu$m thickness). Sample loading was achieved by injecting DFP solution through the remaining open side of the chamber, followed by sealing of the chamber with UV-curable optical adhesive (Norland 63).

\subsection*{Melting curve}
 Samples were prepared by loading unary (B particles) or 1:1 binary (A and B particles) mixtures of particles in the microchamber at a concentration leading to a surface density of approximately 0.01 particles/$\mu m^{2}$ following the gravity sedimentation of the DFPs, owing to the silica density of ca. 2 g/cm$^3$. A Peltier thermoelectric module (TE Technology, Inc) driven by a dipolar thermoelectric temperature controller (TE Technology, Inc) was employed to control the sample temperature. 
 Melting curves were measured by sequentially cooling the DFP samples through the melting transition, with sample equilibration for 15 min at points above and below the transition temperature and for at least 30 min within the transition region at each temperature point. Following equilibration, at least five independent images were collected using an inverted optical microscope (Nikon Eclipse TE2000U, 40x air-immersion objective, 1.5x amplifier). The singlet fraction, defined as the number of individual particles that are not incorporated in a particle aggregate/crystal, was determined by quantitative comparison of the area of individual particles with the area of particles contained in aggregates by common microscopy image analysis\cite{Dreyfus2009, Crocker1996}. 

\subsection*{Colloidal crystallization}
Sample chambers containing suspensions of binary DFP mixtures with tailored particle number ratio, $n$, and ssDNA blending ratios, $\gamma_i$, for $i$=A and B particles were attached to the block of a PCR machine (DNA Thermal Cycler 480, Perkin-Elmer) using silicone grease and incubated with a prescribed temperature trajectory through the melting transition to form crystal structures. Specifically, 
the temperature annealing was initiated above the melting temperature ($\sim$45 $^{o}$C), where most of the particles exist in their singlet state. The sample temperature was reduced at 1 $^{o}$C sequential increments, with each step maintained for $\sim$ 4 hours. The cooling was terminated at room temperature. Resulting two-dimensional DFP structures were observed by fluorescence microscopy (Nikon Eclipse TE2000U) using a 60x oil-immersion objective.

\subsection*{Image processing and pair correlation function}
All image analysis was based on the image processing routines developed by Crocker and Grier\cite{Crocker1996} and code developed by Eric Weeks and implemented in the software package IDL (Exelis Visual Information Solutions). Pair correlation functions with respect to all particles (A and B particles) were calculated from independent brightfield images whereas pair correlation functions with respect, specifically, to green (A) or red (B) particles were calculated from multi-channel fluorescence images.

\subsection*{Molecular dynamics simulation details}
We perform molecular dynamics simulations using the software package LAMMPS~\cite{Plimpton1995} in a canonical ensemble at low packing fraction of 8\%. All the quantities below are reported in LJ units.
The temperature was maintained by a Langevin thermostat with a damping coefficient $\tau$ = 2 
in a 2D simulation box with periodic boundary conditions applied in all (x and y) directions. 
The pair interactions between the particles are modeled using SI eq. 1 to capture interparticle interactions between DFPs. Each simulation is conducted for at least $10^8$ time steps with a step size of $\Delta$t = 0.001. 
The system contains a total of 400 or 402 particles, depending on the particle number ratio, $n$, of A and B particles. 

\subsection*{Identification of the binary crystals formed in MD simulations}
In addition to the nearest neighbor analysis (see SI Fig. S3) and visual inspection, we have implemented the common neighbor analyses (CNA) for the identification of 2D crystals. The python scripting interface of OVITO\cite{Stukowski2009}, originally available for the 3D crystals, was used to calculate the CNA indices - a set of integer triplets (see 
SI Fig. S4 for details) of each particle within the formed crystals. We compare these values with those calculated for reference perfect crystals to distinguish the following binary 2D superlattices: square, hexagonal, alternating string, honeycomb, Kagome and square Kagome. In the determination of CNA indices for the identification of the overall arrangement of DFPs of diameter D$_{p}$, i.e. hexagonal and square lattices, a cut-off radius of R$_{cut}$ = 1.6 D$_{p}$ was used to define the neighborhood, whereas, in the identification of the compositionally ordered crystals, R$_{cut}$ was taken as 1.87 D$_{p}$. The structural identifications were further confirmed by the visual inspection of the grown crystals. On the other hand, the rhombic crystals reported in this work, were only visually identified.


\begin{thebibliography}{10}

\bibitem{jones2015programmable}
M.~R. Jones, N.~C. Seeman, C.~A. Mirkin, {\it Science\/} {\bf 347}, 1260901
  (2015).

\bibitem{Barnaby2015}
S.~N. Barnaby, {\it et~al.\/}, {\it J. Am. Chem. Soc.\/} {\bf 137}, 13566
  (2015).

\bibitem{Ross2015}
M.~B. Ross, J.~C. Ku, M.~G. Blaber, C.~A. Mirkin, G.~C. Schatz, {\it Proc.
  Natl. Acad. Sci.\/} {\bf 112}, 10292 (2015).

\bibitem{Kim2006}
A.~J. Kim, P.~L. Biancaniello, J.~C. Crocker, {\it Langmuir\/} {\bf 22}, 1991
  (2006).

\bibitem{zhang2013general}
C.~Zhang, {\it et~al.\/}, {\it Nat. Mater.\/} {\bf 12}, 741 (2013).

\bibitem{Oh2015}
J.~S. Oh, Y.~Wang, D.~J. Pine, G.~R. Yi, {\it Chem. Mater.\/} {\bf 27}, 8337
  (2015).

\bibitem{Angioletti2012}
S.~Angioletti-Uberti, B.~M. Mognetti, D.~Frenkel, {\it Nat. Mater.\/} {\bf 11},
  518 (2012).

\bibitem{Rogers2015}
W.~B. Rogers, V.~N. Manoharan, {\it Science\/} {\bf 347}, 639 (2015).

\bibitem{Nykypanchuk2008}
D.~Nykypanchuk, M.~M. Maye, D.~van~der Lelie, O.~Gang, {\it Nature\/} {\bf
  451}, 549 (2008).

\bibitem{Macfarlane2011}
R.~J. Macfarlane, {\it et~al.\/}, {\it Science\/} {\bf 334}, 204 (2011).

\bibitem{Li2012}
T.~I. N.~G. Li, R.~Sknepnek, R.~J. MacFarlane, C.~A. Mirkin, M.~{Olvera De La
  Cruz}, {\it Nano Lett.\/} {\bf 12}, 2509 (2012).

\bibitem{Auyeung2014}
E.~Auyeung, {\it et~al.\/}, {\it Nature\/} {\bf 505}, 73 (2014).

\bibitem{Cigler2010}
P.~Cigler, A.~K.~R. Lytton-Jean, D.~G. Anderson, M.~G. Finn, S.~Y. Park, {\it
  Nat. Mater.\/} {\bf 9}, 918 (2010).

\bibitem{Scarlett2011}
R.~T. Scarlett, M.~T. Ung, J.~C. Crocker, T.~Sinno, {\it Soft Matter\/} {\bf
  7}, 1912 (2011).

\bibitem{Casey2012}
M.~T. Casey, {\it et~al.\/}, {\it Nat. Commun.\/} {\bf 3}, 1209 (2012).

\bibitem{Zhang2015}
Y.~Zhang, {\it et~al.\/}, {\it Nat. Mater.\/} {\bf 14}, 840 (2015).

\bibitem{Park2014}
J.~G. Park, {\it et~al.\/}, {\it Angew. Chem. Int. Ed.\/} {\bf 53}, 2899
  (2014).

\bibitem{Gasser2001}
U.~Gasser, E.~R. Weeks, A.~Schofield, P.~N. Pusey, D.~a. Weitz, {\it Science\/}
  {\bf 292}, 258 (2001).

\bibitem{Biancaniello2005}
P.~L. Biancaniello, A.~J. Kim, J.~C. Crocker, {\it Phys. Rev. Lett.\/} {\bf
  94}, 94 (2005).

\bibitem{Dreyfus2010}
R.~Dreyfus, {\it et~al.\/}, {\it Phys. Rev. E\/} {\bf 81}, 1 (2010).

\bibitem{DiMichele2013}
L.~{Di Michele}, {\it et~al.\/}, {\it Nat. Commun.\/} {\bf 4}, 1 (2013).

\bibitem{Wang2015}
Y.~Wang, {\it et~al.\/}, {\it J. Am. Chem. Soc.\/} {\bf 137}, 10760 (2015).

\bibitem{Wang2015i}
Y.~Wang, {\it et~al.\/}, {\it Nat. Commun.\/} {\bf 6}, 7253 (2015).

\bibitem{Dreyfus2009}
R.~Dreyfus, {\it et~al.\/}, {\it Phys. Rev. Lett.\/} {\bf 102}, 5 (2009).

\bibitem{song2016effect}
M.~Song, Y.~Ding, M.~A. Snyder, J.~Mittal, {\it Langmuir\/} {\bf 32}, 10017
  (2016).

\bibitem{Hartmann2002}
D.~M. Hartmann, M.~Heller, S.~C. Esener, D.~Schwartz, G.~Tu, {\it J. Mater.
  Res.\/} {\bf 17}, 473 (2002).

\bibitem{Zou2005}
B.~Zou, B.~Ceyhan, U.~Simon, C.~M. Niemeyer, {\it Adv. Mater.\/} {\bf 17}, 1643
  (2005).

\bibitem{Puchner2008}
E.~M. Puchner, S.~K. Kufer, M.~Strackharn, S.~W. Stahl, H.~E. Gaub, {\it Nano
  Lett.\/} {\bf 8}, 3692 (2008).

\bibitem{Shyr2008}
M.~H.~S. Shyr, D.~P. Wernette, P.~Wiltzius, Y.~Lu, P.~V. Braun, {\it J. Am.
  Chem. Soc.\/} {\bf 130}, 8234 (2008).

\bibitem{kung2014template}
S.-C. Kung, C.-C. Chang, W.~Fan, M.~A. Snyder, {\it Langmuir\/} {\bf 30}, 11802
  (2014).

\bibitem{leunissen2005ionic}
M.~E. Leunissen, {\it et~al.\/}, {\it Nature\/} {\bf 437}, 235 (2005).

\bibitem{rabideau2007computational}
B.~D. Rabideau, R.~T. Bonnecaze, {\it Langmuir\/} {\bf 23}, 10000 (2007).

\bibitem{tkachenko2002morphological}
A.~V. Tkachenko, {\it Phys. Rev. Lett.\/} {\bf 89}, 148303 (2002).

\bibitem{martinez2011design}
F.~J. Martinez-Veracoechea, B.~M. Mladek, A.~V. Tkachenko, D.~Frenkel, {\it
  Phys. Rev. Lett.\/} {\bf 107}, 045902 (2011).

\bibitem{mahynski2015grafted}
N.~A. Mahynski, A.~Z. Panagiotopoulos, {\it J. Chem. Phys.\/} {\bf 142}, 074901
  (2015).

\bibitem{Stukowski2012}
A.~Stukowski, {\it Model. Simul. Mater. Sci. Eng.\/} {\bf 20}, 045021 (2012).

\bibitem{Humphrey1996}
W.~Humphrey, A.~Dalke, K.~Schulten, {VMD: Visual molecular dynamics} (1996).

\bibitem{ferraro2014graphoepitaxy}
M.~E. Ferraro, R.~T. Bonnecaze, T.~M. Truskett, {\it Phys. Rev. Lett.\/} {\bf
  113}, 085503 (2014).

\bibitem{casey2012driving}
M.~T. Casey, {\it et~al.\/}, {\it Nat. Commun.\/} {\bf 3}, 1209 (2012).

\bibitem{jenkins2014hydrodynamics}
I.~C. Jenkins, M.~T. Casey, J.~T. McGinley, J.~C. Crocker, T.~Sinno, {\it Proc.
  Natl. Acad. Sci.\/} {\bf 111}, 4803 (2014).

\bibitem{vo2015stoichiometric}
T.~Vo, {\it et~al.\/}, {\it Proc. Natl. Acad. Sci.\/} {\bf 112}, 4982 (2015).

\bibitem{Steinberg2004}
G.~Steinberg, K.~Stromsborg, L.~Thomas, D.~Barker, C.~Zhao, {\it Biopolymers\/}
  {\bf 73}, 597 (2004).

\bibitem{Crocker1996}
J.~Crocker, D.~Grier, {\it J. Colloid Interface Sci.\/} {\bf 179}, 298 (1996).

\bibitem{Plimpton1995}
S.~Plimpton, {\it J. Comput. Phys.\/} {\bf 117}, 1 (1995).

\bibitem{Stukowski2009}
A.~Stukowski, {\it Modell. Simul. Mater. Sci. Eng.\/} {\bf 18}, 015012 (2009).

\end{thebibliography}
\end{document}


\title{Supplementary Material  for\\
        ``Binary superlattice design by controlling DNA-mediated interactions"}

\author
{Minseok Song, Yajun Ding, Hasan Zerze, Mark A. Snyder, Jeetain Mittal$^{\ast}$\\
\\
\normalsize{Department of Chemical and Biomolecular Engineering, Lehigh University, Bethlehem, PA 18015, USA}\\
\normalsize{$^\ast$To whom correspondence should be addressed; E-mail:  jeetain@lehigh.edu.}
}



\section{Supporting information text}
\subsection{Effective pair potential model for DNA-functionalized particles}

The effective pairwise interactions between DNA-functionalized particles (DFPs) are derived from previously published simulation data using a sequence-specific coarse-grained model~\cite{Ding2014}. Advanced sampling techniques (replica exchange molecular dynamics and umbrella sampling) were used to obtain free energy or potential of mean force (PMF) as a function of interparticle distance. We use the following expression to fit these data to represent various contributions to the interparticle interactions between DFPs, (i) repulsive interactions between particle cores, (ii) repulsive interactions due to DNA chain overlap, and (iii) attractive interactions due to DNA hybridization.

\begin{equation}
U(r)=\epsilon_{o}\left(\frac{\sigma_{o}}{r-r_{shift}}\right)^{n}+\frac{A_{0}}{1+\text{exp}({A_{1}(r-A_{2}))}}-\frac{B_{0}}{1+\text{exp}({B_{1}(r-B_{2}))}}
\end{equation}

Figure S\ref{pmfs} shows an example comparison between the simulation data and eq. 1. By varying $B_{0}$, we can model potentials with different attractive well depths if particle size and overall grafting density is held constant. Even though the PMFs represented by eq. 1 are reflective of nanosized particles, the qualitative shape of the potential is quite similar to the one measured experimentally for micron-sized particles~\cite{Rogers2011}. We simply rescale the potential to reflect features of micron-sized particles with an appropriate DNA length used in the experimental part of this work. Specifically, we have used the following values for various parameters in eq. 1: $\epsilon_{0}$ = 10.0$\epsilon$, $\sigma_{0}$ = 0.2$\sigma$, r$_{shift}$ = 0.8$\sigma$, n = 36, A$_{0}$ = 11.03$\epsilon$, A$_{1}$ = 404.4$\sigma^{-1}$, A$_{2}$ = 1.0174$\sigma$, B$_{1}$ = 1044.5$\sigma^{-1}$ and B$_{2}$ = 1.031$\sigma$. B$_{0}$ was changed from 0 to 1.32 to vary the attractive well depth between 0 and 1.0$\epsilon$, respectively. 

\subsection{Nearest neighbor analysis (NNA)}
To distinguish the formation of various binary superlattices, we calculate the average number of like (AA and BB) and unlike contacts (AB), defined as N$_{K}$ = $\sum n_{K}^{i} / n$ with K = \{AB, AA, BB\}. 
Where, n is the total number of particles analyzed in a given lattice and $n_{K}^{i}$ is the  number of contacts (AB, AA or BB) between particle $i$ and its nearest neighbors. 



\newpage
\section{Supporting Information Figures}

\setcounter{figure}{0}
\begin{figure}[htbp]
\centering
\includegraphics[width=0.5\columnwidth]{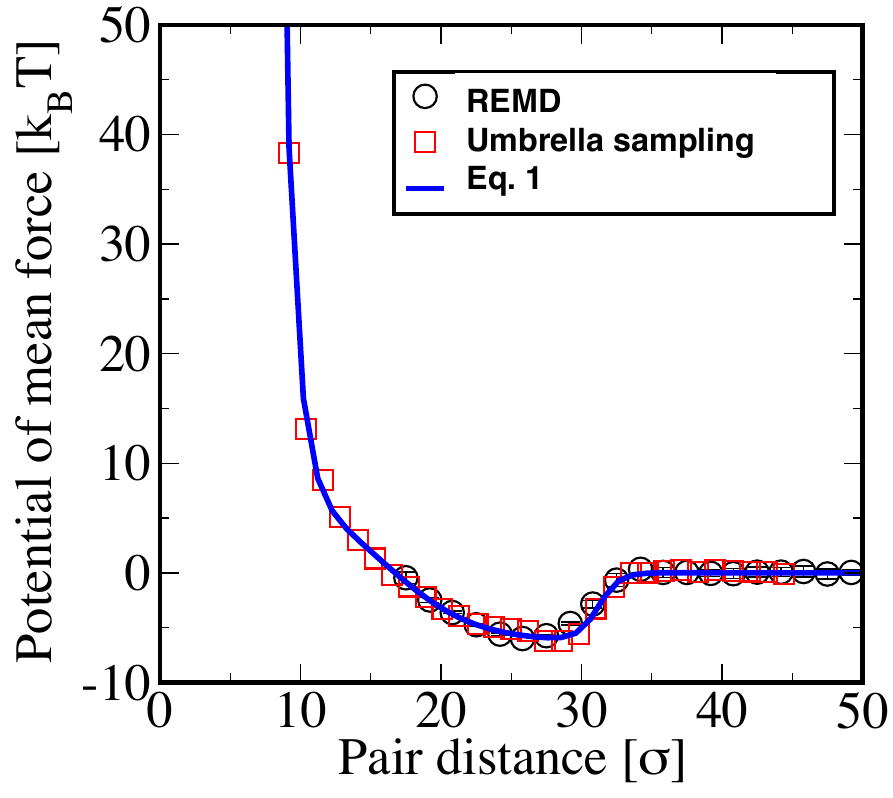}
\renewcommand{\figurename}{Figure S}
\caption{Potential of mean force (PMF) between two DNA-functionalized particles, with each particle grafted by 16 ssDNA strands with sequence TTTTTTATGTATCAAGGT or TTTTTTACCTTGATACAT. 
}
\label{pmfs}
\end{figure}


\begin{figure}[htbp]
\centering
\includegraphics[width=0.45\columnwidth]{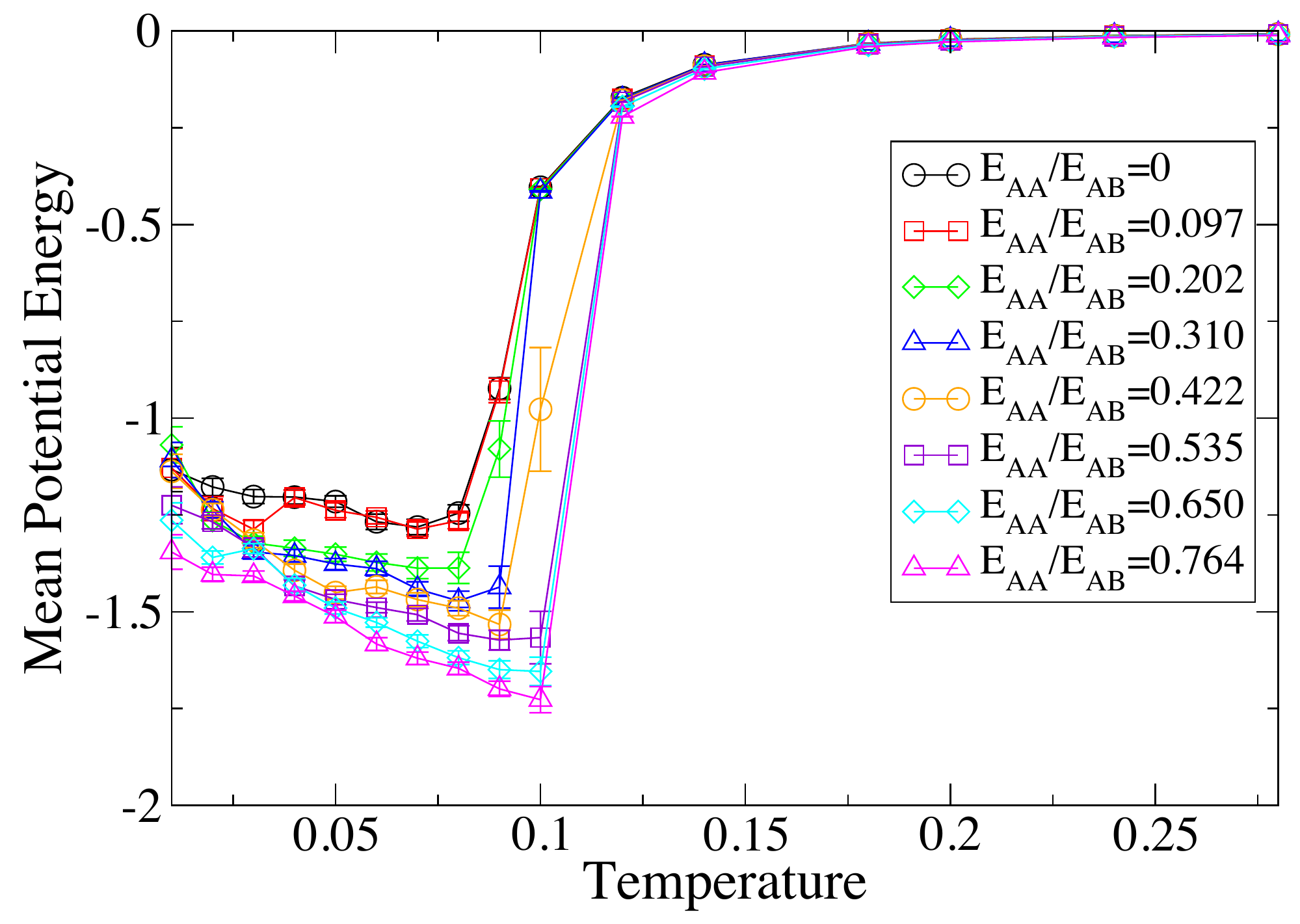}
\renewcommand{\figurename}{Figure S}
\caption{Mean potential energy as a function of temperature from molecular dynamics simulation for different $E_{\mathrm {AA}}/E_{\mathrm {AB}} = E_{\mathrm {BB}}/E_{\mathrm {AB}}$. These data are used as a guide to define putative melting temperature and to conduct additional simulations to study self-assembly behavior of binary superlattices.}
\label{meltsim}
\end{figure}


\begin{figure}[htbp]
\centering
\includegraphics[width=0.6\columnwidth]{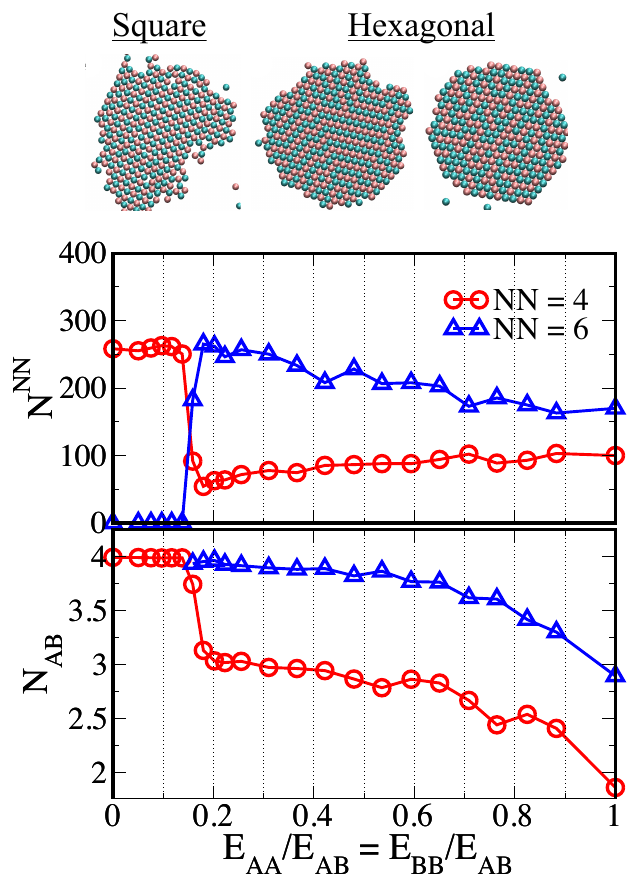}
\renewcommand{\figurename}{Figure S}
\caption{Nearest neighbor analysis (NNA). (Top) Average number of particles with nearest neighbors (NNs) equal to 4 and 6. (Bottom) Average number of unlike contacts ($N_{\mathrm {AB}}$) as a function of $E_{\mathrm {AA}}/E_{\mathrm {AB}} = E_{\mathrm {BB}}/E_{\mathrm {AB}}$. For low pair energies, square lattices (NN = 4) are observed with near-perfect compositional order ($N_{\mathrm {AB}} = 4$), whereas hexagonal lattices (NN = 6) are observed at intermediate and high pair energies. The compositional order for hexagonal lattices also depend on the interparticle interaction strength.  
}
\label{nn4nn6}
\end{figure}




\begin{figure}[htbp]
\centering
\includegraphics[width=\columnwidth]{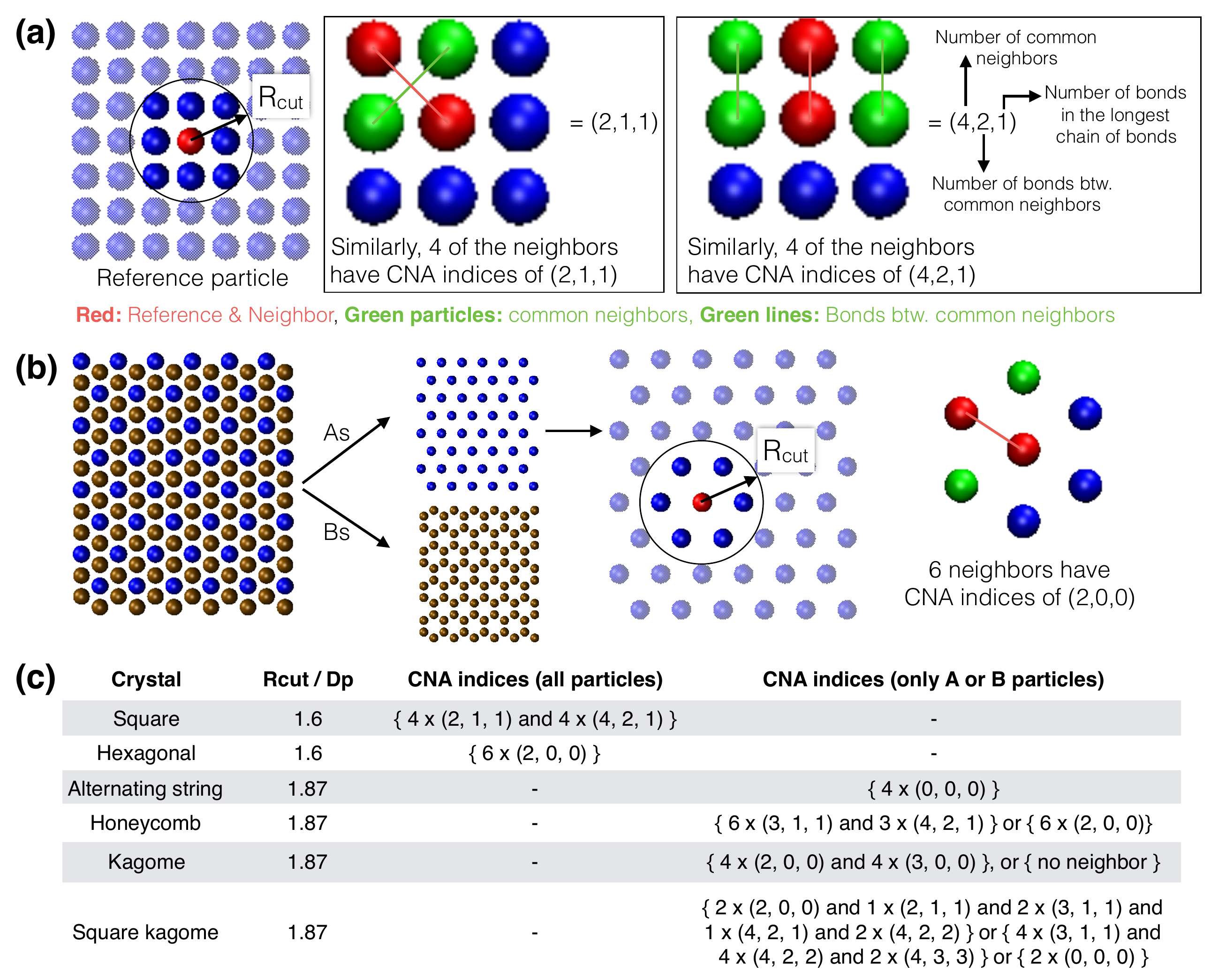}
\renewcommand{\figurename}{Figure S}
\caption{Common neighbor analyses (CNA). (a) Illustration of the determination of CNA indices for a perfect square lattice and the definition of index components. (b) The CNA index determination in a compositionally ordered lattice (illustrated in a perfect honeycomb) for a specific particle type to be used in the identification of the compositionally ordered 2D crystals. (c) Reference CNA indices and frequencies for perfect 2D crystals.}
\label{cna}
\end{figure}






\begin{figure}[htbp]
\centering
\includegraphics[width=0.6\columnwidth]{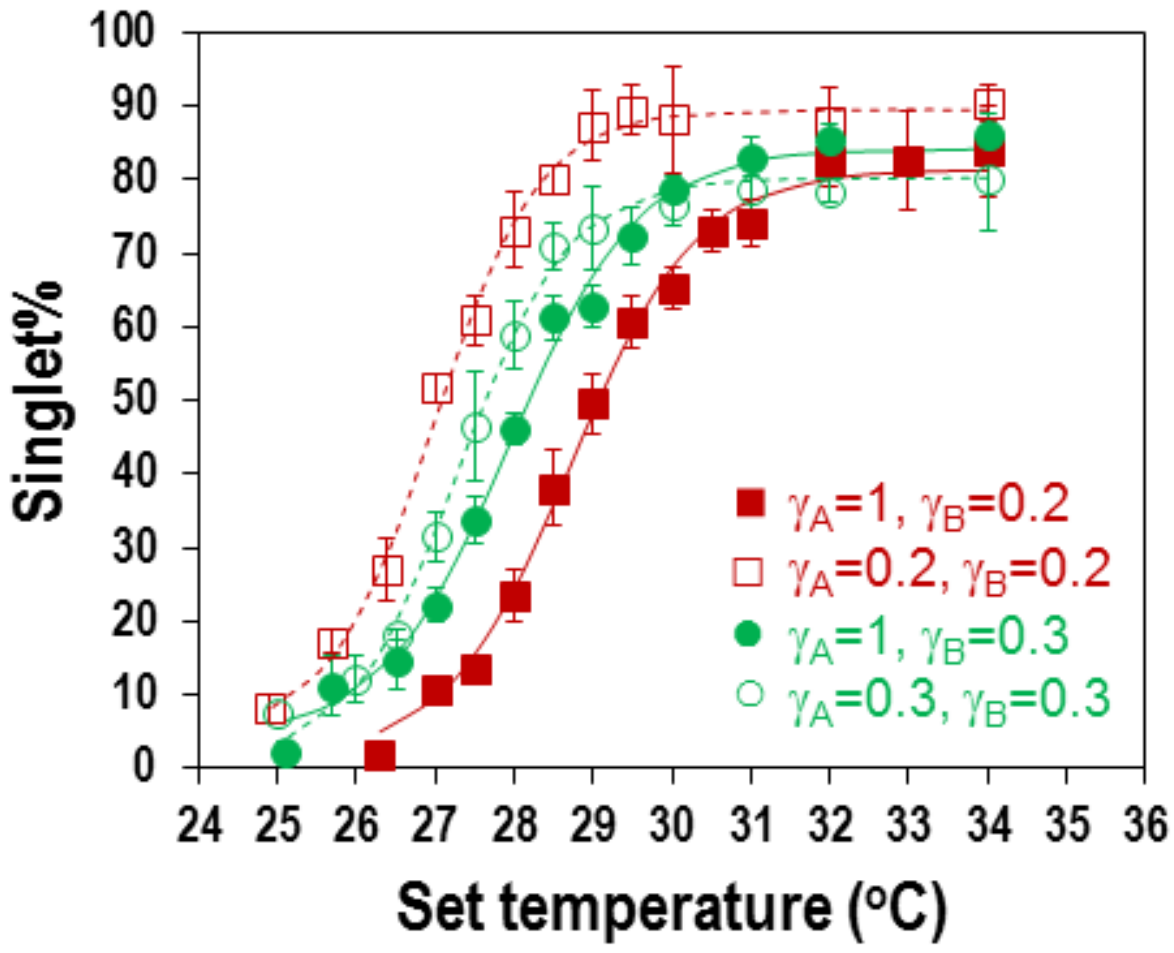}
\renewcommand{\figurename}{Figure S}
\caption{Experimental melting profiles for a suspension of the DNA-functionalized particles. When the fraction of $\alpha$-strands on B-type particles is increased, the melting curve of the binary system ($\gamma_A \ne \gamma_B$) shifts to left whereas that of the unary system ($\gamma_A = \gamma_B$) shifts to the right.}
\label{meltexp}
\end{figure}

\newpage